\documentclass[twocolumn,10pt]{article} 

\usepackage[square,numbers,sort&compress,comma]{natbib}

\usepackage{amsmath}
\usepackage{amssymb}
\usepackage{caption}
\usepackage{graphicx}
\usepackage{latexsym}
\usepackage[all]{nowidow}
\usepackage{comment}
\usepackage{bm}
\usepackage{times}

\usepackage{hyperref}
\usepackage{caption}
\usepackage{subcaption}
\usepackage{xcolor}
\newcommand{\be}{\begin{equation}}
\newcommand{\ee}{\end{equation}}

\newcommand{\ov}{\overline}

\newcommand{\inner}[2]{\big< #1 ,  #2 \big>}

\def\ov{\overline}

\def\exa{\expandafter}

\newcommand{\calZ}{\cal Z}
\newcommand{\calX}{\cal X}

\topmargin - 12pt 
\renewenvironment{abstract}%
              {
               \small
               {\bfseries \abstractname}
               \par
               \vspace{10pt}
              }

\renewcommand\abstractname{Abstract}

\newcommand{\nomenclature}
              [1]
              {
               \bgroup
               \flushleft
               \small\bf
               #1
               \par
               \egroup
              }

\renewcommand{\section}
              [1]
              {
               \bgroup
               \flushleft
               \small\bf
               \stepcounter{section}
               \arabic{section}. #1
               \par
               \egroup
              }

\renewcommand{\subsection}
              [1]
              {
               \bgroup
               \flushleft
               \small\em
               \stepcounter{subsection}
               \arabic{section}.
               \arabic{subsection}. #1
               \par
               \egroup
              }

\renewcommand{\subsubsection}
              [1]
              {
               \bgroup
               \flushleft
               \small\em
               \stepcounter{subsubsection}
               \arabic{section}.
               \arabic{subsection}.
               \arabic{subsubsection}. #1
               \par
               \egroup
              }

  \newcommand{\acknowledgement}
              [1]
              {
               \bgroup
               \flushleft
               \small\bf
               #1
               \par
               \egroup
              }

  \newcommand{\sectionbib}
              [1]
              {
               \bgroup
               \flushleft
               \small\bf
               #1
               \par
               \egroup
              }

\begin{document}

\title{\LARGE Reduced Order Modeling of Turbulence-Chemistry Interactions using Dynamically Bi-Orthonormal Decomposition}

\author{{\large Aidyn Aitzhan$^{a,*}$, Arash G. Nouri$^{a}$, Peyman Givi$^{a}$, Hessam Babaee$^{a}$}\\[10pt]
        {\footnotesize \em $^a$Department of Mechanical Engineering and Materials Science, University of Pittsburgh, Pittsburgh, PA, 15261, USA}\\[-5pt]}

\date{}


\small
\baselineskip 10pt


\twocolumn[\begin{@twocolumnfalse}
\vspace{50pt}
\maketitle
\vspace{40pt}
\rule{\textwidth}{0.5pt}
\begin{abstract} 
%
The performance  of the dynamically bi-orthogonal (DBO) decomposition  for the reduced order modeling  of turbulence-chemistry interactions is assessed.  DBO is an \textit{on-the-fly} low-rank approximation technique, in which the instantaneous composition matrix of the reactive flow field is decomposed into a set of  orthonormal spatial modes, a set of  orthonormal vectors in the composition space, and  a factorization of the low-rank correlation  matrix. Two  factors which distinguish between DBO and the  reduced order models (ROMs) based on the principal component analysis (PCA) are: (i)  DBO does not require any offline data generation; and (ii)  in DBO the low-rank composition subspace is time-dependent as opposed to static subspaces in PCA. Because of these features, DBO can adapt on-the-fly  to intrinsic and externally excited  transient changes in state of the transport variables. 
For demonstration, simulations are conducted of a non-premixed CO/H\textsubscript{2} flame in a temporally evolving jet. The GRI-Mech 3.0 model with 53 species is used for chemical kinetics modeling. The results are appraised  via {\it a posteriori} comparisons against data generated via full-rank direct numerical simulation (DNS)  of the same flame, and the PCA reduction of the DNS data.  The  DBO also yields  excellent predictions of various statistics of the thermo-chemical variables. 
\end{abstract}
\vspace{10pt}
\parbox{1.0\textwidth}{\footnotesize {\em Keywords:} Reduced order modeling, time-dependent basis, dynamically bi-orthonormal decomposition.}
\rule{\textwidth}{0.5pt}
\vspace{10pt}
\end{@twocolumnfalse}] 


\clearpage 

\section{Introduction} \addvspace{10pt}
One of the most  significant challenges with turbulent combustion modeling and simulation  is associated with the high dimensionality of the transport variables as required for its physical description \cite{Bilger2005,Pope13,Raman2019}.  Hydrocarbon combustion, for example, involves from 50 to 7000 species depending on the fuel \cite{LL09,Westbrook2005Computational}. Even with the aid of exascale computing, high-fidelity simulations  of turbulent reactive flows with detailed kinetic remain computationally prohibitive \cite{Chen2011,NGL19}.  To deal with this issue, developments of {\it reduced ordered models} (ROMs) have been at the heart of turbulent reacting flow research for many decades now \cite{LW94,Pope13}. Most of the current methods  are for development of low order manifolds by which all
other transport variables can be calculated.  

Consider the species mass fractions field $\bm{\Phi}(x,t)\equiv \big[\phi_1(x,t) \big| \phi_2(x,t) \big| \dots \big| \phi_{n_s}(x,t) \big]$ for $n_s$ total number of species, with $(x,t)$ denoting the space-time coordinates. The earliest, the simplest, and still one of the most popular models is the {\sl flame sheet approximation} in which the statistics of all of the thermo-chemical variables are related to those of a single conserved variable: the {\sl mixture fraction} \cite{HWH49,Toor62,LW80} (${\calZ} (x,t)$).  The same is the case for flows under chemical equilibrium: $\phi_\alpha (x,t)= \phi_\alpha ({\calZ} (x,t))$.  The simplest extension is  the {\sl Laminar Flamelet Model} (LFM), in which another variable is augmented to the mixture fraction. $\phi_\alpha (x,t)= \phi_\alpha ({\calZ} (x,t), {\calX} (x,t))$, where the choice of the auxiliary ${\calX} (x,t)$ variable depends on the flame structure.  The model originally due to Peters \cite{Peters84}, has been  one of the most popular means of turbulent combustion simulation for several decades now.  For some examples, see Refs.\ \cite{MMG94,DRKC98,PS00,MLP03,Pitsch06,DSMG07,IP08,NYGSP10,Bray16,VDDBTD16,TKHP16}. This popularity is partially due to the model's simplicity.  However, the joint statistics of the mixture fraction and the auxiliary variable cannot be specified in a systematic manner \cite{Peters00}.  

Developments of higher order manifolds via reduced order kinetics have been the subjects of  broad investigations  by a variety of methods \cite{LL09,Pope13}.  In these methods, essentially the composition variables are partitioned into a set of represented (reduced) compositions, and the remaining unrepresented ones. The  manifold is constructed as a  multivariate function of a small number of the  reduced compositions.  The  construction is  either based on the original thermo-chemistry equations with imposition of certain assumptions about their physical-chemical characteristics, or empirically by  analysis of the data  generated via detailed chemistry for a given flow.   Both of these approaches have experienced success.  However, they are all optimized for {\sl specific} flow conditions or  configurations.  It would be desired to have the  manifold reduction technique developed  {\it on-the-fly} for general applications.

Exploiting various correlations via dimensionality reduction is one way to reduce the computational burden of solving turbulent reactive transport equations. See Refs.\ \cite{SP09,OE17} for the application of principal component analysis (PCA) in turbulent combustion. The dimension reduction techniques exploit the correlated structure in the data and they seek the \emph{intrinsic dimensionality} by transforming  the high-dimensional data into a meaningful representation of reduced dimensionality \cite{MPH07,C15}. While  dimensionality reduction techniques are widely used for analysis and data compression  \cite{VMS05,C98,SSM98,L03,TSL00,RS00,LL06,KBBJ16}, or to build ROMs \cite{AIAA17}, they share one important feature: They extract the low-rank structure from \emph{data} and they are designed for a target problem. To reduce the number of species on-the-fly, one approach is to extract these structures from the \emph{model}, \textit{i.e.}  reactive species transport equations and bypass the need to generate data. Recently, some of these techniques, which we refer to as \emph{model-driven} dimension reduction techniques, have  shown promising results for the solution of stochastic partial differential equations (PDEs). Dynamically orthogonal (DO) \cite{SL09,Babaee:2017aa,B19},  bi-orthogonal (BO)    \cite{CHZI13} and dynamically bi-orthonormal decompositions (DBO) \cite{PB20,RNB21}   are  model-driven stochastic ROM techniques. For deterministic linear dynamical systems,  optimally time-dependent (OTD) reduction was introduced \cite{Babaee_PRSA,NBGCL22}.  In all of these model-driven techniques, closed-form (partial) differential  equations for the evolution of the low rank structures (modes) are derived. Leveraging the mathematical rigour of these techniques, for some cases,  numerical and dynamical system analyses are used to shed light into their performance and dynamics \cite{CSK14,MNZ15,BFHS17}. Closely related techniques  can also be found in quantum physics and quantum chemistry  \cite{Beck:2000aa,KL07}.

The  objective of this work  is to develop and implement our   model-driven  strategy \cite{RNB21} to systematically construct  reduced order modeling of turbulence-chemistry interactions in a canonical flow.  The  developments  is not  under any restrictive flow assumption, but is   in conjunction with direct numerical simulation (DNS) in which the lower dimension subspace of the representative fields are  constructed on-the-fly.

\section{Methodology} \addvspace{10pt}
\subsection{Formulation} \addvspace{10pt}
For a variable density reacting flow involving $n_s$ species, the primary transport variables are the fluid density $\rho(x, t) $, the velocity vector  $\bm{v}(x, t)$, the specific enthalpy $h (x, t)$, the pressure $p(x, t)$, and the species mass fractions $\phi_{\alpha}(x, t) \
(\alpha=1,2,\dots,n_s)$. The conservation equations governing the transport variables are the continuity, momentum, enthalpy (energy) and species mass fraction equations, along with an equation of state in the format detailed in Ref.\ \cite{Aitzhan2021arXiv}. The transport equations for composition space (species mass fractions and enthalpy) can be stated in the matrix form as:
\begin{equation}\label{eq:compos_trans}
\frac{\partial \Phi}{\partial t} = - (\bm{v} \cdot \nabla )\Phi + \frac{1}{\rho}  \nabla \cdot ( \nabla \Phi \Gamma)+ S(\rho,\Phi),
\end{equation}
where $\Phi(x,t) \in \mathbb{R}^{\infty \times n_c}$ is the composition matrix with $n_c$ columns $(n_c=n_s+1)$, and  $\infty$ denotes the continuous direction $(x)$:
\begin{align}
\Phi(x,t) =[\phi_1(x,t) &\mid \phi_2(x,t) \mid \dots \nonumber \\  \dots &\mid \phi_{n_s}(x,t) \mid  \phi_{n_c}(x,t)],
\end{align}
in which $\phi_{n_c} \equiv h$. In Eq.\ (\ref{eq:compos_trans}), $\Gamma\in \mathbb{R}^{n_c \times n_c}=\mbox{diag}\{ \gamma_1, \gamma_2, \dots, \gamma_{n_c}\}$ is the diagonal matrix of diffusivities at each spatial point, and $S$ denotes the chemical source term. The inner product in the spatial domain between two fields $f(x)$ and $g(x)$ is defined as:
\begin{equation*}\label{eq:Inner Product}
\inner{f(x)}{g(x)} = \int_{D} f(x)g(x)dx,
\end{equation*}
where $D$ denoted the physical domain, and the  $L_2$ norm induced by this inner product as:
\begin{equation*}\label{eq:L2}
\big\|f(x)\big\|_2 = \inner{f(x)}{f(x)}^{\frac{1}{2}}.
\end{equation*}
The Frobenius norm of a matrix $A(x)=[a_1(x), a_2(x), \dots, a_m(x) ] \in \mathbb{R}^{\infty \times m}$ is defined as:
\begin{equation*}\label{eq:LF}
    \big\|A(x)\big\|_{\mathcal{F}} ^2= \sum_{i=1}^m \int_{D} a_i^2(x) dx.
\end{equation*}
A column-wise inner product between two matrices $A(x,t)\in\mathbb{R}^{\infty\times m}$ and $B(x,t)\in \mathbb{R}^{\infty\times n}$ is defined as 
\begin{equation*}
    Q(t) = \inner{A(x,t)}{B(x,t)},
\end{equation*}
where $Q(t)\in\mathbb{R}^{m\times n}$ is a matrix with components $Q_{ij}(t) = \inner{a_i(x,t)}{b_j(x,t)}$, where $b_j(x,t)$ denotes the $j$th column of $B(x,t)$. Therefore,  $\inner{AR_A}{B}=R_A^T\inner{A}{B}$ and $\inner{A}{BR_B}=\inner{A}{B}R_B$ for any $R_A \in \mathbb{R}^{m \times m}$ and $R_B \in \mathbb{R}^{n \times n}$. 

The goal is to solve for a low-rank decomposition of  $\Phi(x,t)$ instead of solving Eq.\ (\ref{eq:compos_trans}). To this end, the DBO decomposition for the full species concentration matrix is considered:
\begin{equation}\label{eq:DBO}
\Phi(x,t) = \sum_{j=1}^{r} \sum_{i=1}^{r} u_i(x,t)\Sigma_{ij}(t)y^T_j(t) + E(x,t),
\end{equation}
 where
\begin{subequations}\label{eq:USigmaY}
\begin{align}
U(x,t) &= [u_1(x,t)\mid u_2(x,t)\mid ...\mid u_r(x,t)],\\
Y(t) &= [y_1(t)\mid y_2(t)\mid ...\mid y_r(t)].
\end{align}
\end{subequations}
Here, $r<n_c$ is the reduction size, $U(x,t)\in \mathbb{R}^{\infty \times r}$ is a matrix whose columns are a set of time-dependent orthonormal spatial modes, $\Sigma(t)\in \mathbb{R}^{r\times r}$  is a factorization of the reduced   correlation matrix, $Y(t)\in \mathbb{R}^{n_c \times r}$ is the matrix of  time-dependent orthonormal composition  modes, and $E(x,t) \in \mathbb{R}^{\infty \times n_c}$ is the reduction error. The key observation regarding  Eq.~(\ref{eq:DBO}) is that all three components are time dependent. This enables the decomposition to adapt on-the-fly to the changes in $\Phi(x,t)$. In fact, the DBO decomposition closely approximates the instantaneous singular value decomposition (SVD) of $\Phi(x,t)$.  Orthonormality of spatial modes and  composition modes implies that:
\begin{subequations}\label{eq:Orthonormality}
\begin{align}
\inner{u_i(x,t)}{u_j(x,t)} =\delta _{ij}, \label{eq:U-Ortho}\\ 
y^T_i(t)y_j(t) =\delta _{ij}  \label{eq:Y-Ortho},
\end{align}
\end{subequations}
where $\delta_{ij}$ is the Kronecker delta. The variational principle whose optimality conditions lead to  closed form evolution equations for the components of the DBO decomposition is given by:
\begin{align}\label{eq:VarPrin}
\mathcal{F} (\dot{U}, \dot{\Sigma}, \dot{Y}) =  \left \|  \frac{\partial}{\partial t}[U\Sigma Y^T] - \mathcal{M}(U\Sigma Y^T)  \right \|_{\mathcal{F}}^2,
\end{align}
subject to the orthonormality conditions as  given by Eqs.\ (\ref{eq:U-Ortho})-(\ref{eq:Y-Ortho}). 
In Eq.~(\ref{eq:VarPrin}),  $\dot{( ~ )} \equiv \partial ( ~ )/\partial t$ and $\mathcal{M}$ is the right hand side of composition transport equation: $\mathcal{M} =  -(\bm{v} \cdot \nabla )U\Sigma Y^T + \frac{1}{\rho}  \nabla \cdot ( \nabla U\Sigma Y^T \Gamma)+ S(\rho,U\Sigma Y^T)$.  The variational principle given in Eq.\ (\ref{eq:VarPrin}) seeks to minimize the difference between the right hand side of the species transport equation  and the time derivative of the DBO decomposition subject to the orthonormality constraints given by Eqs.\ (\ref{eq:U-Ortho})-(\ref{eq:Y-Ortho}).  The control parameters are $\dot{U}$, $\dot{Y}$ and $\dot{\Sigma}$. As derived in Ref.\ \cite{RNB21}, the closed-form evolution equations of $U(x,t)$, $\Sigma(t)$ and $Y(t)$ are: 
\begin{subequations}
\label{eq:evol_USY}
\begin{align}
\frac{\partial U}{\partial t}  &=   (\mathcal{M}Y - U\inner{U}{\mathcal{M}Y})\Sigma^{-1}, \label{eq:evol_U}\\
\frac{d \Sigma}{d t} &= \inner{U}{ \mathcal{M}Y}, \label{eq:evol_S} \\
\frac{dY}{dt}  &= (I-Y Y^T) \inner{\mathcal{M}}{U}\Sigma^{-T}. \label{eq:evol_Y}
\end{align}
\end{subequations}

\subsection{Computational Cost} \addvspace{10pt}
The main computational advantage of using DBO is to evolve only $r$ spatial modes (Eq. (\ref{eq:evol_U})) instead of $n_c$ composition transport equations (Eq.\ (\ref{eq:compos_trans})). The computational cost of evolving  $\Sigma$ and $Y$ is negligible as they are governed by low-rank ordinary differential equations (ODEs). Moreover, in the DBO decomposition, the compositions are stored in the \emph{compressed form}, \textit{i.e.}, matrices $U$, $\Sigma$ and $Y$ are kept in the memory as opposed to their multiplication $U\Sigma Y^T$, \emph{i.e.,} the \emph{decompressed form}.  The memory storage requirement is dominated by $U$ as $\Sigma$ and $Y$ are low-rank matrices and their storage cost is negligible. Therefore,  in comparison to the full species transport equation, this results in the memory compression ratio of $n_c/r$. Reference \cite{RNB21} discusses the efficient way of calculating each terms on the right-hand-side of Eq.\ (\ref{eq:evol_USY}).

\subsection{Comparison against DNS \& PCA} \addvspace{10pt}
 The canonical representation of DBO is used for comparison against DNS and PCA. In this representation, the spatial and composition modes are ranked based on their ``energy" in the second-norm sense. The ranking can be achieved by performing  SVD of $\Sigma(t)$:
\begin{equation}
    \Sigma(t) = R_U(t)\tilde{\Sigma}(t)R_Y(t),
\end{equation}
where $\tilde{\Sigma}(t)$ is a diagonal matrix that contains the ranked singular values: $\tilde{\sigma}_1(t) > \tilde{\sigma}_2(t) >...> \tilde{\sigma}_r(t)$ and $R_U(t)$ and $R_Y(t)$ are orthonormal matrices that can be used to rotate $U$ and $Y$ as follows:
\begin{subequations}\label{eq:RankedUY}
\begin{align}
\tilde{U}(x,t) &= U(x,t)R_U(t), \label{RankedU}\\
\tilde{Y}(t) &= Y(t)R_Y(t). \label{RankedY}
\end{align}
\end{subequations}
The components $\{\tilde{U}(x,t), \tilde{\Sigma}(t),\tilde{Y}(t)\}$ represent the DBO decomposition in the canonical form. It is noted that the DBO in the canonical form, and the form that is computed are equivalent: $U\Sigma Y^T = \tilde{U}\tilde{\Sigma} \tilde{Y}^T$.

The results via DBO are compared with those generated by the reduction based on the instantaneous principal component analysis (I-PCA). The I-PCA components are calculated in a data-driven manner (based on DNS data) by computing SVD of the instantaneous matrix of full species: $\Phi(x,t)=\hat{U}(x,t)\hat{\Sigma}(t)\hat{Y}^T(t)$, where $\hat{( ~ )}$ denotes the components of I-PCA,  $\hat{U}(x,t)=\{\hat{u}_1(x,t), \hat{u}_2(x,t), \dots, \hat{u}_{n_c}(x,t)\}$ is the matrix of left singular functions, $\hat{\Sigma}(t) = \mbox{diag}(\hat{\sigma}_1(t),\hat{\sigma}_2(t), \dots, \hat{\sigma}_{n_c}(t))$ is the diagonal matrix of singular values and $\hat{Y}(t) = \{\hat{y}_1(t),\hat{y}_2(t), \dots, \hat{y}_{n_c}(t) \}$ is the matrix of right singular vectors.

The composition modes of PCA are static, and are derived by the SVD of the whole composition data from DNS through all time steps (not instantaneously like I-PCA) and spatial locations. 
\begin{subequations}
\label{eq:DBO-PCA}
\begin{align}
\mbox{PCA:} \quad \Phi(x,t) &\simeq  U_{PCA}(x,t)Y_{PCA}^T,\label{eq:PCA} \\ 
\mbox{DBO:} \quad \Phi(x,t) &\simeq  U(x,t)\Sigma (t)Y(t)^T\label{eq:DBO-M}.
\end{align}
\end{subequations}
Equation\ (\ref{eq:DBO-PCA}) shows the contrast  between time-dependent subspace extracted by  DBO,  and the static manifolds extracted from PCA. It should be emphasized that $Y_{PCA}$ is comparable with $Y(t)\Sigma(t)$ from DBO. Please refer to \cite{RNB21} for more details.

\section{Flow Configuration and Model Specifications} \addvspace{10pt} 

\begin{figure}
    \centering
    \includegraphics[width=0.48\textwidth]{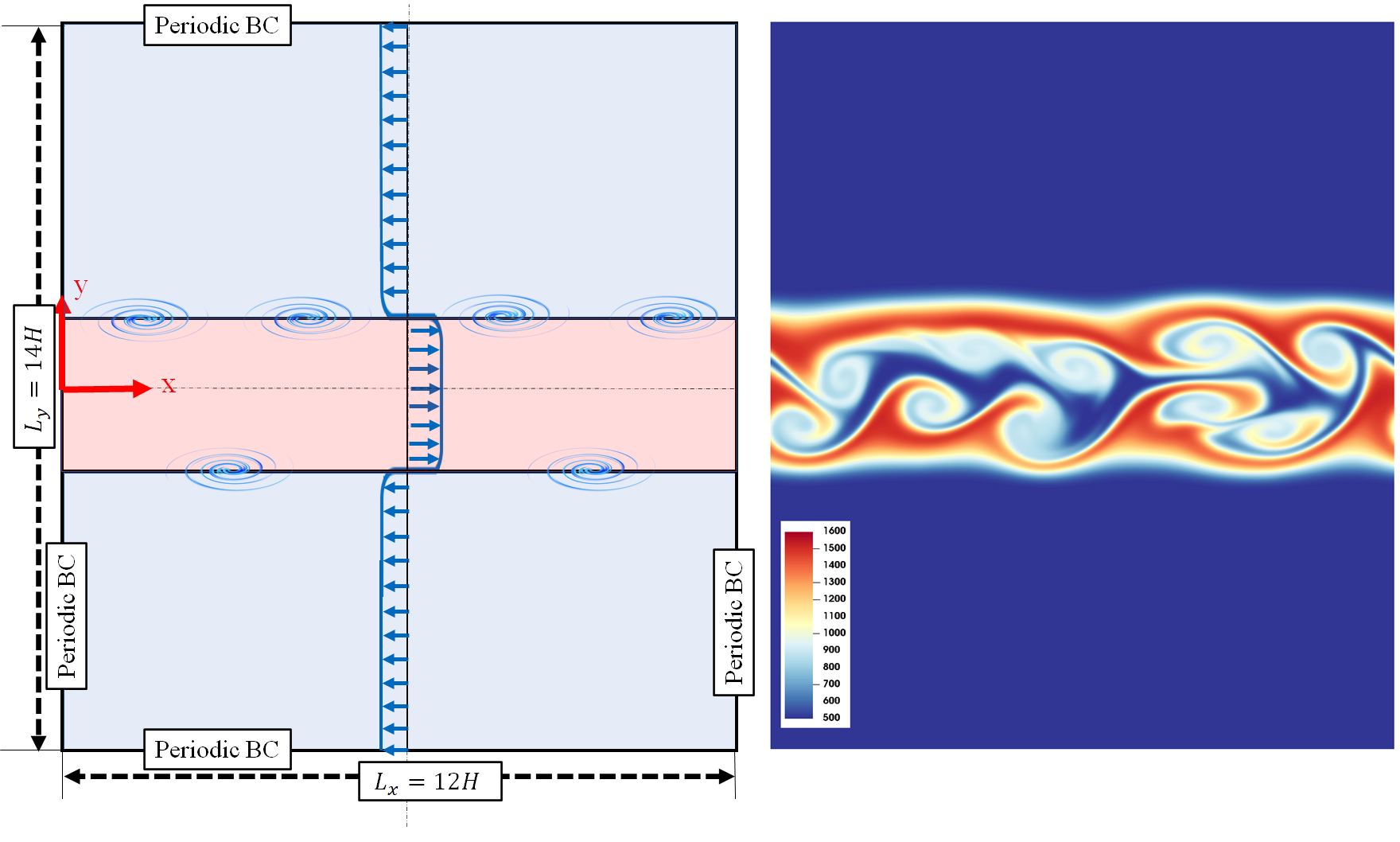}
    \caption{Schematics of the temporally developing turbulent jet flame.}
    \label{fig:FLOW}
\end{figure}

For demonstration, the canonical configuration of a temporally developing planar CO/H\textsubscript{2} jet flame is considered. This flame has been the subject of previous detailed DNS \cite{Hawkes2007Scalar}, and several subsequent modeling and simulations \cite{YANG2013Large, Punati2011Evaluation, Vo2018MMC, Yang2017Sensitivity, Sen2010Large, Aitzhan2021arXiv}. The flame is rich with strong flame-turbulence interactions resulting in local extinction followed by re-ignition. The configuration as considered here is the two-dimensional version of that  in Ref.\ \cite{Hawkes2007Scalar} and is depicted in Fig.\ \ref{fig:FLOW}. The jet consists of a central fuel stream of width $H=0.72$mm surrounded by counter-flowing oxidizer streams. The fuel stream is comprised of 50$\%$ of CO, 10$\%$ H\textsubscript{2} and 40$\%$ N\textsubscript{2} by volume, while oxidizer streams contain 75$\%$ N\textsubscript{2} and 25$\%$ O\textsubscript{2}. The initial temperature of both streams is 500K and thermodynamic pressure is set to 1 atm. The velocity difference between the two streams is $U = 145$m/s. The fuel stream velocity and the oxidizer stream velocity are $U/2$ and $-U/2$, respectively. The initial conditions for the velocity components and mixture fraction are taken directly from center-plane DNS in Ref.\ \cite{Hawkes2007Scalar}, and then the spatial fields of species and temperature are reconstructed from flamelet table generated with $\chi = 0.75 \chi_{\text{crit}}$, where $\chi$ and $\chi_{\text{crit}}$ are scalar dissipation rate and its critical value respectively. The boundary conditions are periodic in stream wise ($x$) and cross-stream wise ($y$) directions. The Reynolds number, based on $U$ and $H$ is $Re=2510$. The sound speeds in the fuel and the oxidizer streams are denoted by $C_1$ and $C_2$, respectively and the Mach number $Ma = U / \left(C_1 + C_2 \right) \approx 0.16$ is small enough to justify a low Mach number approximation. The combustion chemistry is modelled via the GRI-Mech 3.0 mechanism \cite{GRI} containing 53 species with 325 reaction steps. The DNS of the base reactive flow is conducted via the PeleLM solver \cite{Nonaka2018Conservative} as detailed in Ref.\ \cite{Aitzhan2021arXiv}.
Simulations are conducted for the duration $0 \le t \le 60t_j$,  $t_j=H/U$, and the mixture fraction field, denoted by $Z(x,t)$ is constructed from the simulated data \cite{Bilger1976Structure}.

\section{Results and Discussions} 
\begin{figure}
    \centering
    \includegraphics[width=0.48\textwidth]{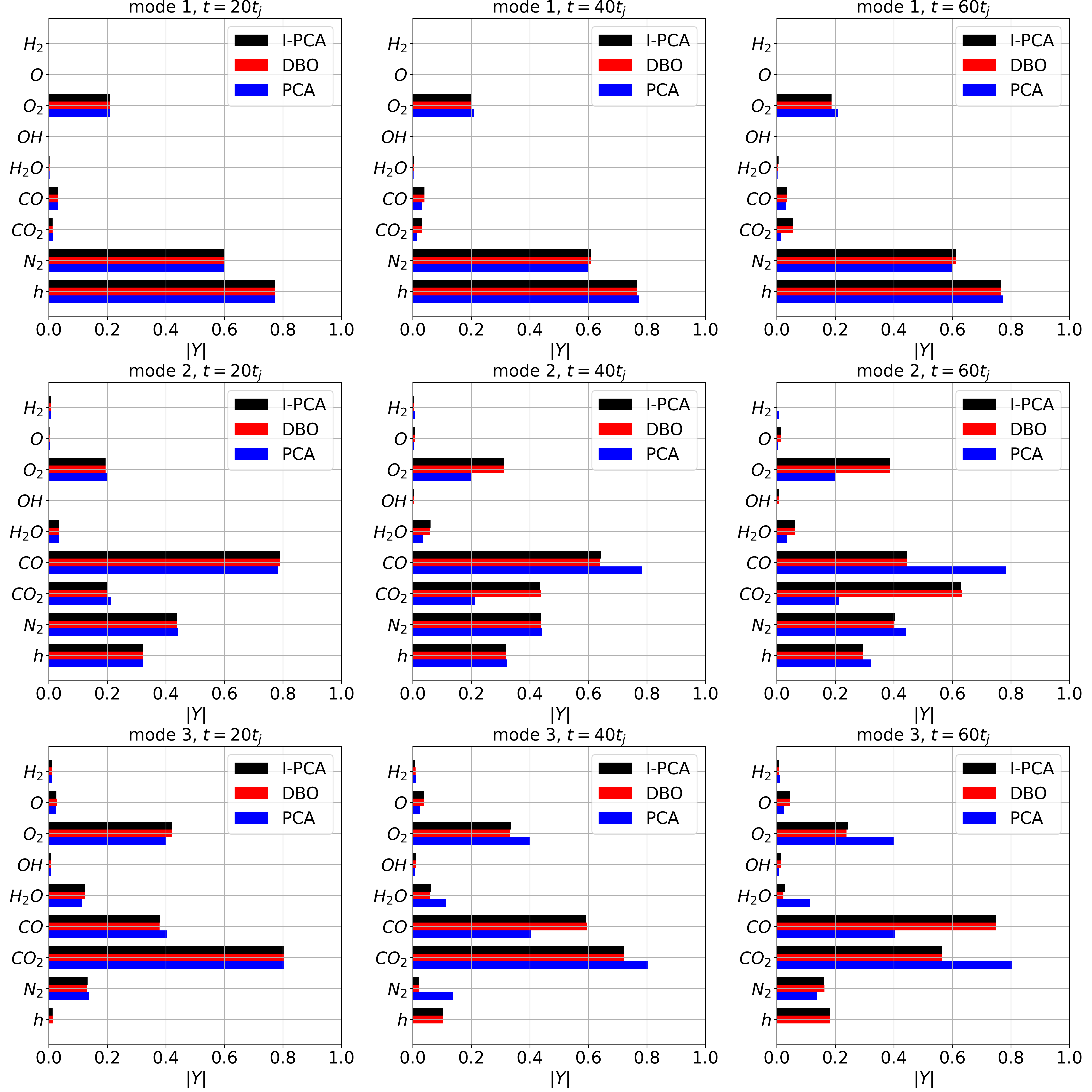}
    \caption{Temporal evolution of species modes $Y$.  For clarity, only most dominant scalars are shown.}
    \label{fig:Ymodes}
\end{figure}

\addvspace{10pt}
Two DBO-ROMs with ranks $r = 6$ and $r = 8$  are considered in conjunction with  GRI-Mech 3.0 mechanism. In Fig.\ \ref{fig:Ymodes}, the low-rank composition subspace is shown for I-PCA, PCA, and DBO.   In particular, the absolute values of the first three dominant modes are shown. The species with very  small $Y$ values are eliminated for clarity. The PCA and I-PCA are both extracted from the DNS data.  It is observed that the PCA subspace is static, but the DBO subspace evolves with time.  The subspace extracted by DBO  matches well with the  low-rank subspace extracted by I-PCA, which is the optimal instantaneous subspace in the $L_2$ sense. The main difference between the PCA  and DBO is that in the former the composition subspace ($Y_{PCA}$) extracts low-rank structures in a \emph{time-averaged} sense. In fact, $Y_{PCA}$ are the dominant eigenvectors of the space- and time-averaged correlation matrix constructed from  the DNS data. The DBO composition subspace can change temporally  and can closely approximate the eigenvectors of the \emph{instantaneous} correlation matrix, as it is evidenced by  a good agreement between $Y_{I-PCA}$ and $Y_{DBO}$.  The  DBO reduction does not require any data, and all of its components ($U$, $\Sigma$ and $Y$) are  extracted directly from the composition transport equations (Eqs.\ (\ref{eq:evol_U}-\ref{eq:evol_Y})).

\begin{figure}
    \centering
    \includegraphics[width=0.45\textwidth]{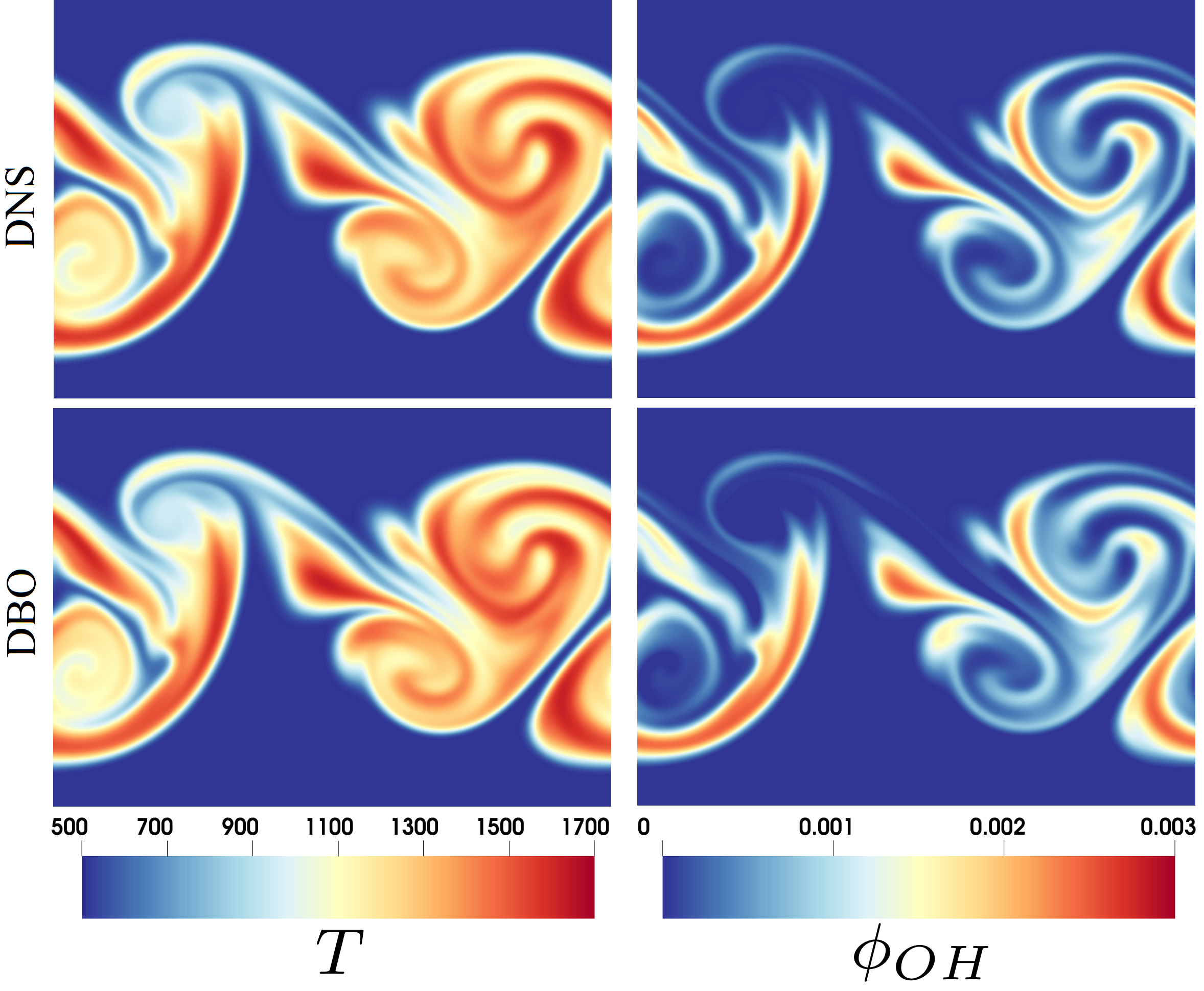}
    \caption{Contour plots of the temperature ($T$) and the hydroxyl radical mass fraction ($\phi_{OH}$) fields at $t=30t_j$ obtained from DNS and DBO with $r=6$.}
    \label{fig:T_DNS_DBO6}
\end{figure}
\begin{figure*}[h!]
\centering
\begin{subfigure}[b]{0.45\textwidth}
   \centering
   \includegraphics[width=\textwidth]{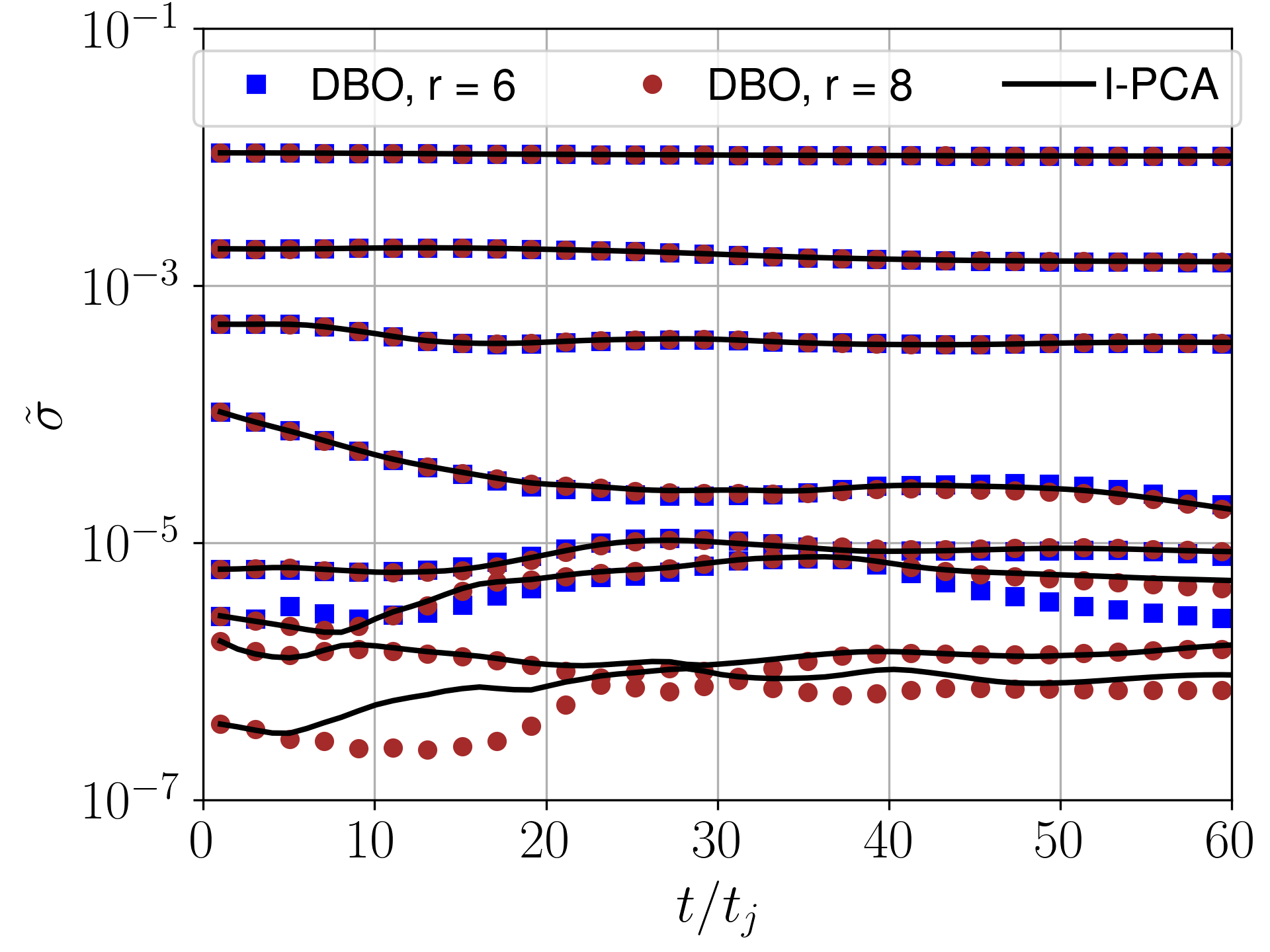}
   
   \vspace{-0.35cm}
   
   \caption{}
   \label{fig:sigma_DBO_GRI}
\end{subfigure}
\begin{subfigure}[b]{0.45\textwidth}
   \centering
   \includegraphics[width=\textwidth]{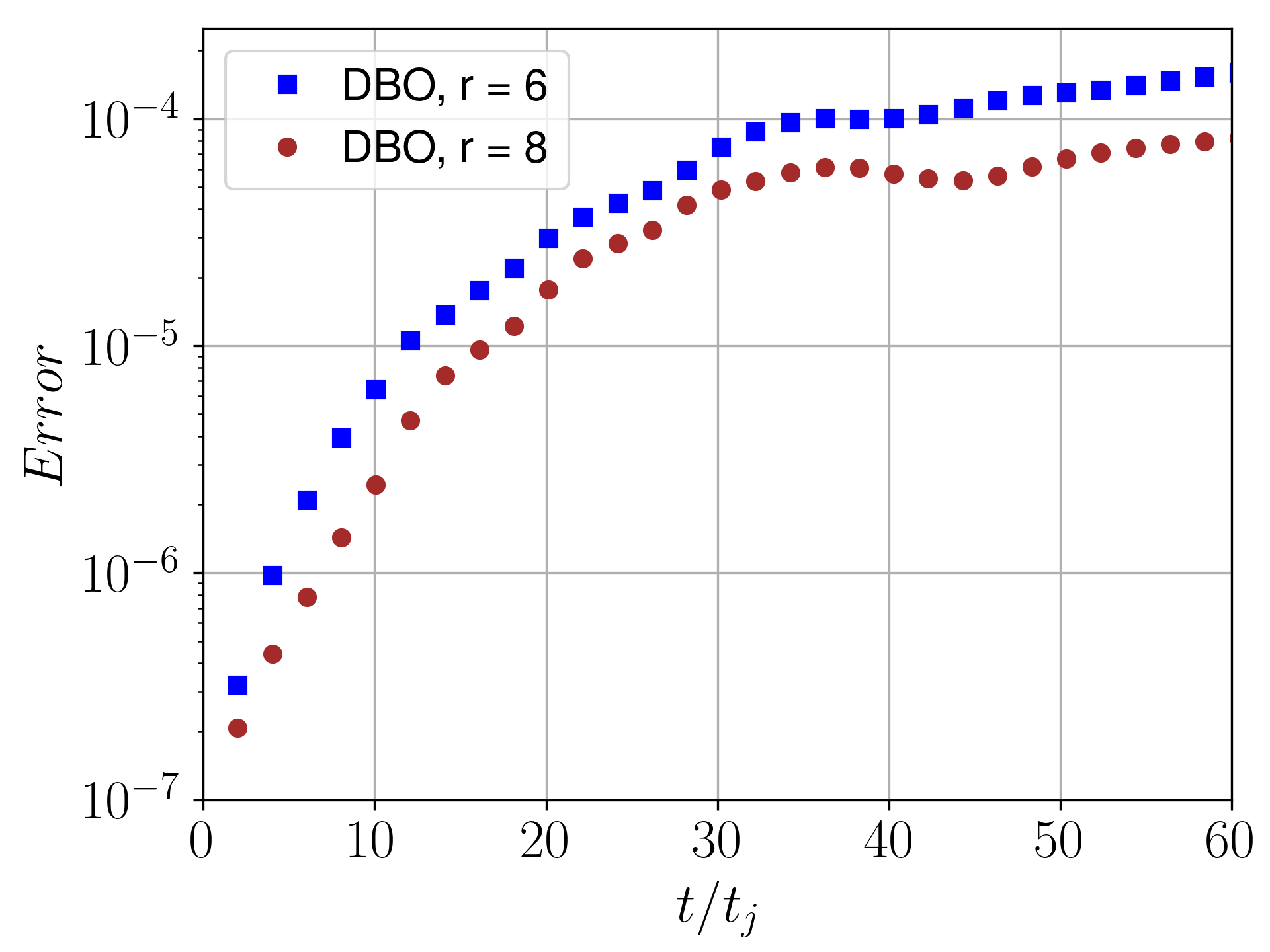}
   
   \vspace{-0.35cm}
   
   \caption{}
   \label{fig:error_DBO_GRI}
\end{subfigure}

\vspace{0.15cm}

\caption{Temporal evolution of (a) singular values and (b) Frobenius error in DBO.}
\label{fig:sigma_error_DBO_GRI}
\end{figure*}

For a visual inspection, in Fig.\ \ref{fig:T_DNS_DBO6}, the temperature ($T$) and the mass fraction of the hydroxyl radical  ($\phi_{OH}$) as obtained by DNS and DBO ($r=6$) are presented.  The agreement is very good.   Figure \ref{fig:sigma_error_DBO_GRI}(a) provides  the comparison between singular values obtained from I-PCA, \textit{i.e.}, the optimal reduction and those obtained from DBO with ranks $r=6$ and $r=8$. For both cases the leading singular values are captured very well. However, there is an error in capturing the lowest singular values due to the effects of the unresolved subspace. This error can be  reduced by increasing the reduction rank $r$ as shown in Fig.\ \ref{fig:sigma_error_DBO_GRI}(b).

\begin{figure}[h!]
    \centering
    \includegraphics[width=0.45\textwidth]{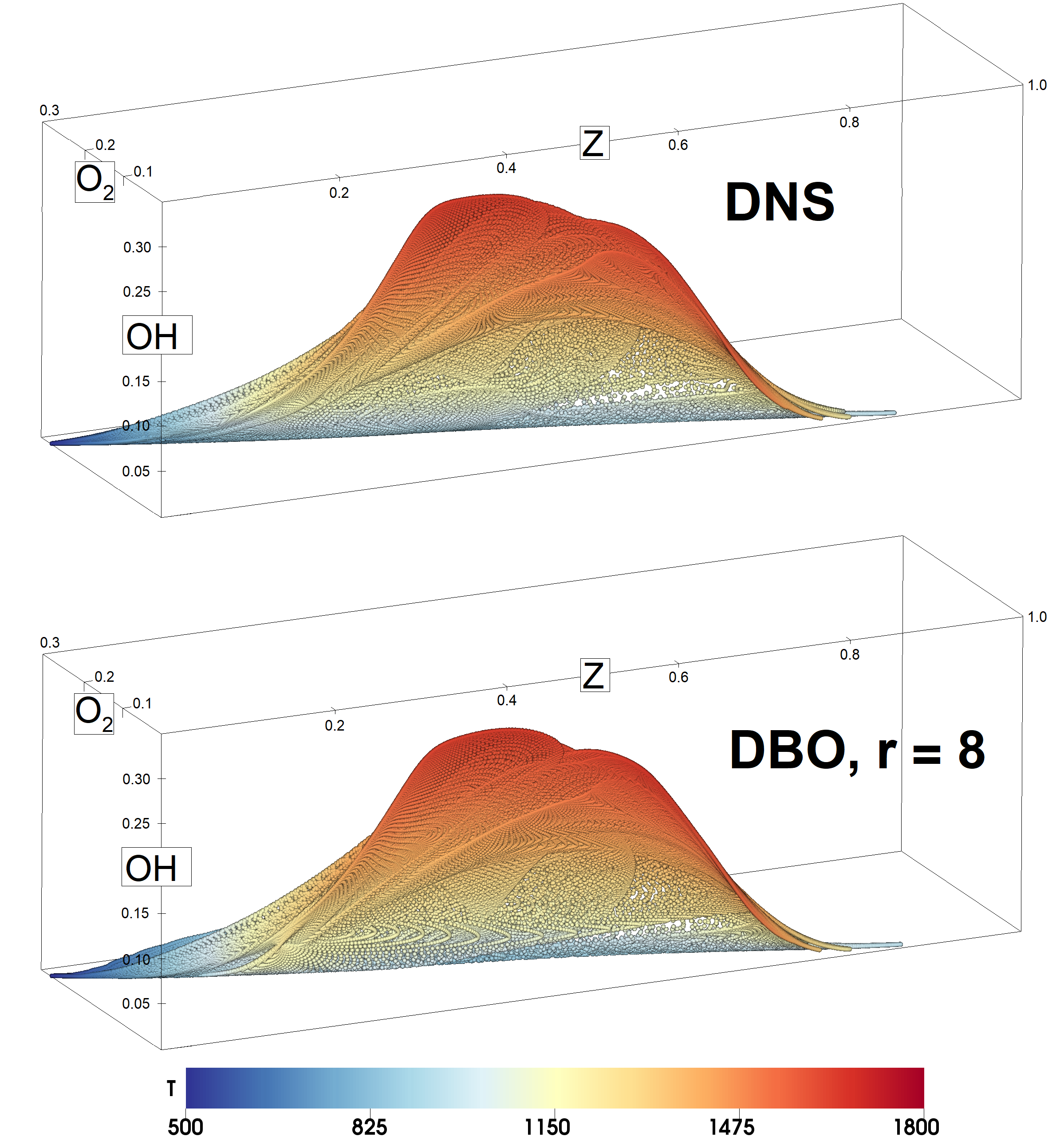}
    \caption{3D manifolds of mixture fraction ($Z$), mass fractions of oxygen ($\phi_{O_2}$) and hydroxyl radical ($\phi_{OH} \times 100$) colored by temperature ($T$) at $t = 40 t_j$.} 
    \label{fig:3D_manifold}
\end{figure}
\begin{figure*}[h!]
    \centering
    \includegraphics[width=0.95\textwidth]{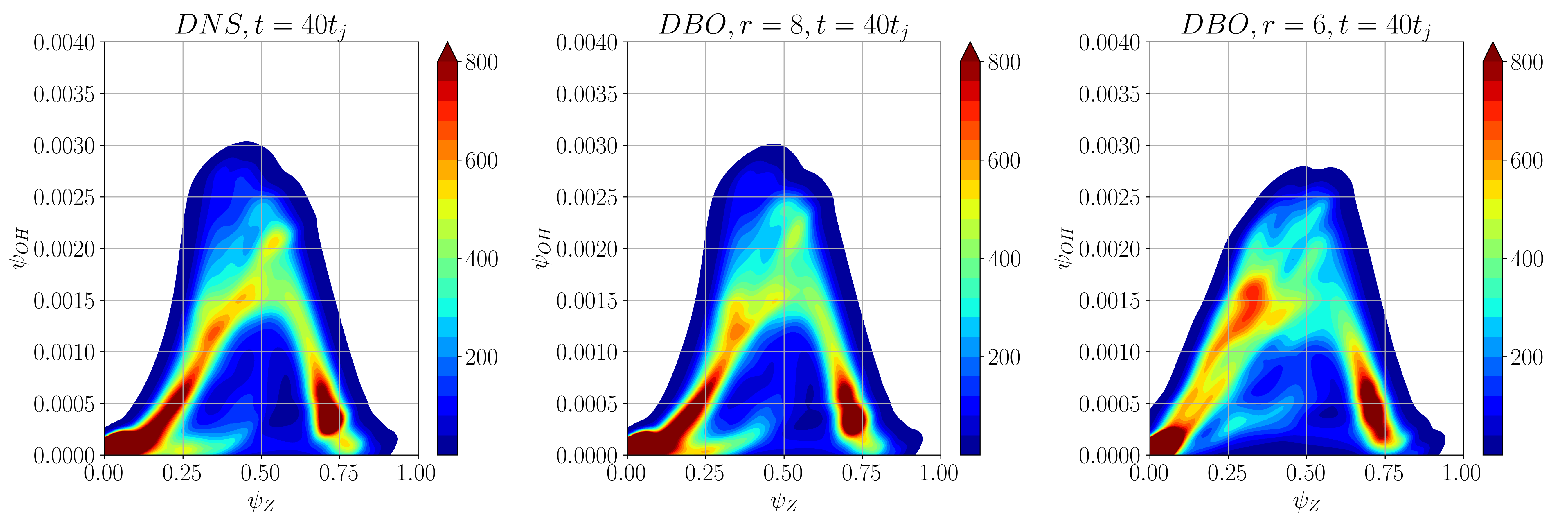}
    \caption{Joint PDFs of mixture fraction and hydroxyl radical, $P \left(\psi_Z, \psi_{OH} \right)$.} 
    \label{fig:jpdf_Z_OH}
\end{figure*}

For a comparative assessment of the  compositional structure of the flame,  representative manifolds encompassed by  O\textsubscript{2}-OH-Z, and temperature are presented in Fig.\  \ref{fig:3D_manifold}.  This figure corroborates the agreements observed  via visual inspections of the flowfield.  Comparisons are also made of various statistics of the reactive trasnport ariables.   The joint probability density functions (PDFs)  of $Z$ and $\phi_{OH}$ ($P \left(\psi_Z, \psi_{OH} \right)$) are generated by data sampled over the entire domain and are presented in Fig.\ \ref{fig:jpdf_Z_OH}. Here $\psi_Q$ denotes the sample space of scalar $Q$. The  agreement between DBO approximation and DNS data are very good and, as expected, improves as $r$ increases.  These PDFs are very similar to those attained in previous three-dimensional DNS of the same fuel \cite{Hawkes2007Scalar}. 

\begin{figure}[h!]
    \centering
    \includegraphics[width=0.48\textwidth]{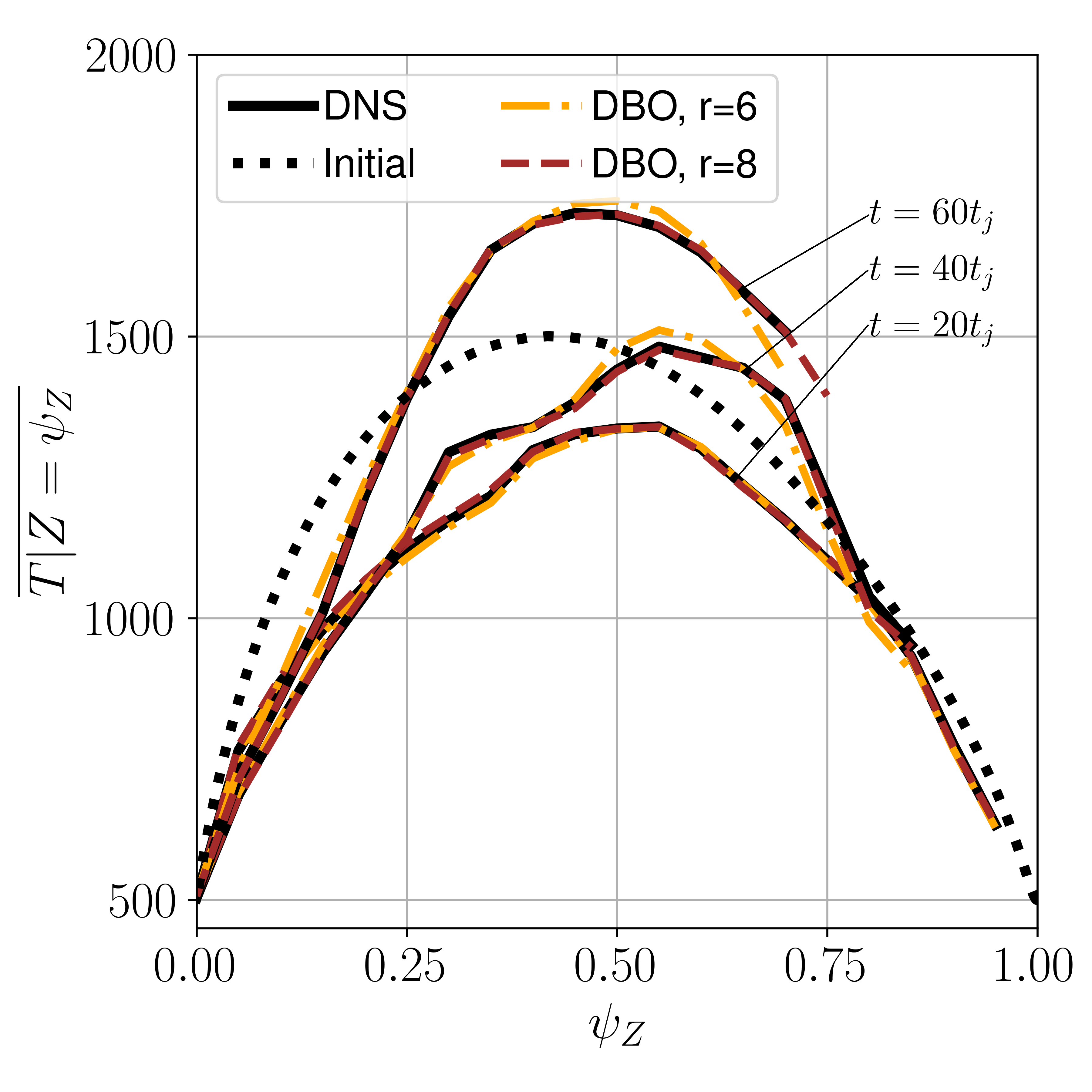}
    \caption{Conditionally averaged temperature values.}
    \label{fig:conditional_T}
\end{figure}

To highlight the non-equilibrium  character of the flame, the   averaged temperature values, conditioned on the instantaneous mixture fraction $\ov{T|Z = \psi_Z}$ are shown in Fig.\ \ref{fig:conditional_T}. Here, the overbar denotes the volume averaging and the vertical bar denotes the conditional statistics. The trends  portrayed in this figure are identical to those in previous DNS \cite{Hawkes2007Scalar} exhibiting initial flame extinction and its subsequent re-ignition.   The excellent agreements between DBO and DNS  is simply due to the fact that bases $U$ and $Y$ adapt to the   physics for the conditions (in this case extinction and re-ignition).  The higher the $r$ value is, the more details DBO resolves and more of the physics is captured.

\section{Conclusions} \addvspace{10pt}
An assessment is made of the performance of a recently developed dynamically bi-orthogonal (DBO) reduced-order modeling of  turbulent combustion \cite{RNB21}. The DBO decomposition is an on-the-fly ROM technique with two distinguishing characteristics in comparison to PCA-based ROMs: (i) in DBO, the low-rank composition subspace is time-dependent as opposed to static; and (ii) no offline data generation is required in DBO and closed-form evolution equations for the low-rank subspaces are derived.   The performance of the DBO approximation is assessed via simulations of a non-premixed CO/H\textsubscript{2} flame in a temporally evolving jet. The GRI-Mech 3.0 model with 53 species is used to model the chemical kinetics. This flame exhibits strong non-equilibrium effects including  extinction and  subsequent re-ignition.  The simulated results clearly indicates that DBO yields  excellent predictions of various statistics of the thermo-chemical variables.  This warrants future implementation of DBO for simulation of complex turbulent combustion systems \cite{GSP12}.


\newpage 

\acknowledgement{Acknowledgments} \addvspace{10pt} 

This work is sponsored by the National Science Foundation (NSF), USA under Grant CBET-2042918, and by the Air Force Office of Scientific Research award (PM: Dr.\ Fariba Fahroo) FA9550-21-1-0247. Computational resources are provided by the Center for Research Computing (CRC) at the University of Pittsburgh.


 \footnotesize
 \baselineskip 9pt


\bibliographystyle{pci_new}
\bibliography{Refs.bib} 


\newpage

\small
\baselineskip 10pt



\end{document}